\documentclass[11pt,draft]{article}
 \usepackage{amsfonts}
 \usepackage{amssymb}
 \def\adt{\dot \alpha}

 \newfont{\bbbold}{msbm10}
 \def\com{\mbox{\bbbold C}}

 \def\cO{{\cal O}}

 \newfont{\goth}{eufm10 scaled \magstep1}

 \def\gl{\mbox{\goth l}}

 \def\gs{\mbox{\goth s}}

 \def\a{\alpha}
 
 \def\c{\gamma}
 \def\d{\delta}

 \def\l{\lambda}\def\L{\Lambda}

 \def\s{\sigma}

 \def\be{\begin{equation}}\def\ee{\end{equation}}
 \def\bea{\begin{eqnarray}}\def\eea{\end{eqnarray}}
 \def\ba{\begin{array}}\def\ea{\end{array}}

 \def\del{\partial}

 \def\xz{\times}

 \def\nab{\nabla}

 \def\del{\partial}
 
 \def\bt{\bullet}

 \let\la=\label

 \let\bm=\bibitem{}

 \def\nn{\nonumber}
 \def\bd{\begin{document}}
 \def\ed{\end{document}}
 \def\bea{\begin{eqnarray}}\def\barr{\begin{array}}\def\earr{\end{array}}
 \def\eea{\end{eqnarray}}
 \def\ft#1#2{{\textstyle{{\scriptstyle #1}\over {\scriptstyle #2}}}}
 \def\fft#1#2{{#1 \over #2}}
 \newcommand{\eq}[1]{(\ref{#1})}
 \def\eqs#1#2{(\ref{#1}-\ref{#2})}
 \def\det{{\rm det\,}}
 \def\tr{{\rm tr}}\def\Tr{{\rm Tr}}

\begin{document}

 \thispagestyle{empty}

 \hfill{KCL-TH-01-??}

  \hfill{\today}

 \vspace{20pt}

 \begin{center}
 {\Large{\bf A note on composite operators in $N=4$ SYM}}
 \vspace{30pt}

 {P.J. Heslop and P.S. Howe} \vskip 1cm {Department of Mathematics}
 \vskip 1cm {King's College, London} \vspace{15pt}

 \vspace{60pt}

 {\bf Abstract}

 \end{center}

 We discuss composite operators in $N=4$ super Yang-Mills theory
 and their
 realisations as superfields on different superspaces. The
 superfields that realise various operators on analytic superspace
 may be different in the free, interacting and quantum theories. In
 particular, in the quantum theory, there is a restricted class of
 operators that can be written as analytic tensor superfields. This
 class includes all series B and C operators in the theory as well
 as some series A operators which saturate the unitarity bounds.
 Operators of this type are expected to be protected from renormalisation.

 {\vfill\leftline{}\vfill \vskip  10pt

 \baselineskip=15pt \pagebreak \setcounter{page}{1}


Over the past few years the Maldacena conjecture \cite{maldacena}
has rekindled interest in four-dimensional superconformal field
theories and this has led to the discovery of many new and
interesting results. Most of these results have concerned
properties of short (series C) operators and their correlation
functions derived both directly in field theory and from
supergravity via the AdS/CFT correspondence. Some recent reviews
and lists of references can be found in \cite{agmo,f,b,hw}. A
striking feature of such operators is that their shortness protects
them from renormalisation - they cannot develop anomalous
dimensions because the representations under which they transform
determine these dimensions uniquely. More recently, however, it has
been found that certain series A operators, which are not short in
the above sense and which had not been anticipated to be protected
from renormalisation, turn out to also have vanishing anomalous
dimensions. These results have been established using the OPE and AdS/CFT
\cite{afp1,afp2}, from partial non-renormalisation of
four-point functions \cite{epss,aeps}, in perturbation theory
\cite{bkrs0} and, most recently, using the OPE in $N=2$ harmonic
superspace \cite{aes}.

The representations of the superconformal group  are well-known
\cite{screp} and their realisations on superfields have been
studied by many authors, see for example
\cite{af1,fz,hh,afsz,fs,hesh}. In particular, shortening conditions for series A representations which saturate unitarity bounds have been discussed in \cite{fz,fs}.
In this note we point out that the
series A operators fall into three distinct classes when looked at
as explicit functions of the underlying $N=4$ supersymmetric
Yang-Mills field strength superfield. There are 3 different types
of behaviour: (i) operators which do not saturate the unitarity
bounds, even in the free theory, (ii) operators which saturate the
unitarity bounds in the free theory but for which the number of
components changes  in the interacting theory and (iii) operators
which saturate unitarity bounds in the interacting theory. This
classification holds in the classical theory where the dimensions
are still (half) integral. In the quantum theory operators of types
(i) and (ii) can develop anomalous dimensions because there are
``nearby'' representations with non-integral quantum numbers which
have the same number of components. On the other hand, for
operators of type (iii) this is not the case, and one therefore
expects them to be protected in a similar fashion to the short
representations of series B and C. All the operators which have
been found to be non-renormalised in references \cite{afp1,afp2,epss,aeps,bkrs0,aes} 
are of type (iii)
as one might expect, but this classification suggests that there
are very many more of them.

Operators of type (i) take care of themselves in that there are no
shortening conditions even in the free case. However, it is not so
easy to distinguish between operators of types (ii) and (iii)
merely by looking at the quantum numbers of the representations or
at their realisation as (abstract) superfields in Minkowski
superspace. It turns out that the operators of type (iii) are those
that can be written as products of chiral primary operators,
possibly with spacetime or spinorial derivatives. Operators of type
(ii) include single trace operators (with the exception of the
chiral primaries) and more complicated operators which include such
single trace functions as factors. The basic reason for this is
that the constraints (on Minkowski superspace) which type (iii)
superfields must satisfy in order to saturate unitarity bounds
follow from the constraints on the gauge-invariant factors whereas,
for operators of type (ii), this is not the case, so that the
corresponding interacting multiplets have more components than 
the free ones. One
way of seeing this is to work on analytic superspace, this having
the advantage that there are no further constraints to be imposed
apart from analyticity. A general  analytic superconformal field
will transform under the isotropy subgroup of the superconformal
group which defines analytic superspace in a non-trivial manner,
i.e. it will have superindices (whereas one can find analytic
superspaces for the series C operators where no indices are
required \cite{fs,hesh}). We shall work on the analytic superspace with the
smallest number of odd coordinates and the smallest number of
additional even coordinates compatible with these. All
representations with (half) integral dimensions can be constructed
from a set of free Maxwell field strength superfields and
derivatives with respect to the coordinates of this analytic
superspace.\footnote{This is briefly discussed in \cite{hesh2}; a
detailed account is in preparation.} The difference between
operators of types (ii) and (iii) can be stated very simply in this
context: operators of type (ii) cannot be so represented in the
interacting case because this would involve applying
gauge-covariant derivatives to the non-Abelian SYM field strength
superfield and this is not allowed because the Yang-Mills potential
is not itself a field on analytic superspace. Operators of type
(iii) are therefore composite operators for which the analytic
superspace derivatives only act on gauge-invariant factors. The
claim, therefore, is that all such series A operators which satisfy
a unitarity bound should be protected from renormalisation. In the
quantum theory operators of types (i) and (ii) both cease to be
realised as analytic tensor superfields. They can still be viewed
as analytic fields but their transformation properties are not of
the usual tensorial type. On the other hand, operators of type
(iii) are analytic tensor superfields even in the quantum theory.
In this sense one can view protection from renormalisation as being
due to analyticity even for series A operators.

Before discussing this in more detail we shall briefly discuss an
example of each type of operator in $N=4$ super Minkowski space.
The field strength superfield $W_I$ transforms under the 6 of
$SO(6)$, and is subject to the constraint

\be
\nab_{\a i} W_I = (\s_I)_{ij} \L_{\a}^j \la{1} \ee

where $\a$ is a 2-component spinor index, $i$ is an $SU(4)$ index
and $\s_I$ is an $SO(6)$ $\s$-matrix. The spinorial derivative
includes a gauge field in the non-Abelian case. The leading
component of $W_I$ is the set of six scalar fields of $N=4$ SYM
while the leading component of $\L_{\a}^i$ is the quartet of spin
one-half fields. The supercurrent is $T_{IJ}=\tr(W_I
W_J)-1/6\,\tr(W_K W_K)$. From \eq{1} it obeys the constraint that
when $D_{\a i}$ is applied to it only the 20-dimensional
representation of $SU(4)$ survives. The quantum numbers specifying
a representation of the $N=4$ superconformal group are
$(L,J_1,J_2,a_1,a_2,a_3)$ where $L$ is the dilation weight, $J_1$
and $J_2$ are spin labels and $(a_1,a_2,a_3)$ are $SU(4)$ Dynkin
labels. We thus see that $T_{IJ}$ has quantum numbers
$(2,0,0,0,2,0)$. The unitarity bounds are:

 \be
 \ba{lll}
 {\rm Series\ A:} & L\geq 2+2J_1 + 2m_1 -{m\over2}\qquad &L\geq
 2+2J_2 +{m\over2} \nn\\
 &&\nn\\
 {\rm Series\ B:} & L= {m\over2};\ L\geq 1+m_1 + J_1,\  J_2=0& {\rm or}\nn \\
&&\nn\\
&L=2m_1-{m\over2}; L\geq 1 + m_1 + J_2, \ J_1=0&\nn\\
&&\nn\\
 {\rm Series\ C:} & L=m_1={m\over2} &J_1=J_2=0
 \ea
 \ee

where $m$ is the total number of boxes in the Young tableau of the
$SU(4)$ representation and $m_1$ the number of boxes in the first
row.

An operator of type (i) is given by $T_{IJ} T_{IJ}$. This has
quantum numbers $(4,0,0,0,0,0)$. It is a series A operator which
does not saturate either unitarity bound and is simply an
unconstrained scalar superfield on Minkowski superspace. In the
quantum theory there is nothing to prevent this operator developing
an anomalous dimension.

An example of a type (ii) operator is the $N=4$ Konishi multiplet,
$K=\tr(W_I W_I)$ \cite{k,hst}. In the free theory this operator
obeys the constraint

\be
D_{ij}K:= D_{\a i} D^{\a}_j K=0 \la{2} \ee

However, in the interacting theory one finds

\be
D_{ij}K\sim \tr([W_{ik}, W_{jl}] W^{kl}):= S_{ij} \la{3} \ee

so that $K$ is now an unconstrained superfield
($W_{ij}:=(\s_I)_{ij} W_I$). This is similar in some respects to
the behaviour of the Yang-Mills supercurrent in ten dimensions. In
the free theory this consists of a quasi-superconformal multiplet
($128+128$) together with a constrained scalar superfield
\cite{bdr} whereas in the interacting theory the scalar superfield
is unconstrained \cite{hnvp}. As in the type (i) case, in the
quantum theory, there is nothing to stop $K$ developing an
anomalous dimension and it is well-known that this indeed happens
\cite{agj, bkrs}.

For an example of type (iii) we consider the operator
$\cO_{IJ}:=T_{IK} T_{JK} -1/6\,\d_{IJ} T_{KL} T_{KL}$, which
transforms under the $20'$ representation of $SU(4)$. This has
quantum numbers $(4,0,0,0,2,0)$; it is a series A operator which
saturates both unitarity bounds. This operator obeys the same
constraints in the interacting theory as it does in the free theory
because they can be derived from the gauge-invariant constraints
that $T_{IJ}$ satisfies. There is a representation related to this
one by changing $L=4$ to $L=4+2\c$ where $\c$ is a real number, but
it has many more components and so one expects $\cO_{IJ}$ to be
protected from renormalisation. Indeed, this operator is one of
those found to have vanishing anomalous dimensions in references
\cite{afp1,afp2,epss,aeps,bkrs0,aes}.

To discuss these operators more generally we shall use super Dynkin
diagrams. For the (complexified) superconformal group $SL(4|N)$
acting on $\com^{4|N}$, the Dynkin diagram depends on the choice of
basis. If the basis is ordered in the standard fashion, 4 even - N
odd, we have the distinguished basis with one odd root, but we
shall use a different basis, which we shall refer to as physical,
in which the basis has the ordering, 2 even - N odd - 2 even. The
physical basis has two odd roots so that the Dynkin diagram is
\be
\begin{picture}(220,20)(-10,-10)
\put(0,0){\makebox[0pt][l]{$\bt\hspace{1.5em}\ominus\hspace{1.5em}
    \underbrace{\bt \hspace{1.5em}
\bt\hspace{1.5em}\cdots
\hspace{1.5em}\bt\hspace{1.5em}\bt}_{N-1}\hspace{1.5em}\ominus
\hspace{1.5em}
\bt$} \rule[.5ex]{6.85em}{.1ex} $\hspace{4.5em}$
\rule[.5ex]{6.6em}{.1ex}
 }
\end {picture}
\la{4} \ee Any representation can be specified by giving labels
associated to each node of the Dynkin diagram. The labels
associated with the two external even (black) nodes are determined
by the spin quantum numbers $(J_1,J_2)$ and the $(N-1)$ internal
even labels are fixed by the Dynkin labels of $SL(N)$. The two odd
(white) labels are then determined by the dilation ($L$) and the
R-symmetry ($R$) quantum numbers. All the Dynkin labels should be
non-negative integers except for the odd ones which can be positive
real numbers. These continuous labels are directly related to
anomalous dimensions of operators.

The super Dynkin diagram can also be used to represent coset spaces
determined by parabolic subgroups. With respect to a given basis
the Borel subalgebra consists of lower triangular matrices, and a
parabolic subalgebra (which by definition is one which contains the
Borel subalgebra) consists of lower block triangular matrices. The
size of these blocks is determined by a set of at most $N+3$
positive integers $k_1 < k_2 \dots$ and can be represented on the
Dynkin diagram by placing crosses through the $k_i$th nodes
(starting from the left). For example, super Minkowski space is
represented by
\be
\begin{picture}(220,0)(-5,0)
\put(0,0){\makebox[0pt][l]{$\bt\hspace{1.5em}\otimes\hspace{1.5em}
\bt\hspace{1.5em}
\bt\hspace{1.5em}\cdots
\hspace{1.5em}\bt\hspace{1.5em}\bt\hspace{1.5em}\otimes
\hspace{1.5em} \bt$} \rule[.5ex]{6.7em}{.1ex} $\hspace{4.5em}$
\rule[.5ex]{6.9em}{.1ex}
 }
\end {picture}
\la{5} \ee

Chiral superspaces have a single cross through one of the odd
nodes, harmonic superspaces have crosses through both odd nodes and
some internal nodes, and analytic superspaces have crosses only
through internal nodes. Superspaces with crosses through the
external nodes include projective super twistor space, but such
spaces are inconvenient for representation theory and so will not
be considered further here.

The crosses on a super Dynkin diagram factorise the diagram into
sub-(super)-Dynkin diagrams corresponding to the semi-simple
subalgebra of the Levi subalgebra (the diagonal blocks in the
parabolic), while the Dynkin labels above the crosses correspond to
charges under internal $U(1)$'s or dilation and R weights. In
general the Levi subalgebra will be a superalgebra and so the
fields can carry superindices. Only in cases where both odd nodes
have crosses through (such as for super Minkowski space and
harmonic superspaces) does the Levi subalgebra contain no
superalgebra.

In order to have unitary representations (of the real
superconformal group $SU(2,2|N)$) the Dynkin labels on the odd
nodes must exceed those of the adjacent external nodes by at least
one unless one or both pairs of these adjacent nodes are zero. This
gives three series of unitarity bounds. We label the nodes from the
left $n_1 \dots n_{N+3}$ so that the two odd nodes are $n_2$ and
$n_{N+2}$ and the adjacent external nodes are $n_1$ and $n_{N+3}$
respectively. For series A we have $n_2 \geq n_1 +1$ and $n_{N+3}
\geq n_{N+2}+1$. For series B we have either $n_1=n_2=0$ and
$n_{N+3} \geq n_{N+2}+1$ or we have $n_2 \geq n_1 +1$ and $n_{N+3}
= n_{N+2}=0$. Finally series C requires that $n_1=n_2=n_{N+3} =
n_{N+2}=0$. For general $N$ we have

\bea n_2&=&{1\over2}(L-R) + J_1 + {m\over N} - m_1 \nn\\
n_{N+2}&=&{1\over2}(L+R) + J_2 -{m\over N} \la{6} \eea

where $m$ is the total number of boxes in the internal Young
tableau determined by the $SU(N)$ Dynkin labels $(a_1,\ldots
a_{N-1})=(n_3,\ldots n_{N+1})$ and $m_1$ is the number of boxes in
the first row. The external black labels are
$(n_1,n_{N+3})=(2J_1,2J_2)$. For $N=4$ we need to impose $R=0$ in
order to have representations of $PSU(2,2|4)$.

The above discussion implies that all of the unitary
representations can be represented in various ways on superfields
defined on superspaces, and that these fields will transform
linearly under representations of the Levi subalgebra. In
particular, in $N=4$, all of the representations can be realised as
(analytic) superfields on $(N,p,q)=(4,2,2)$ analytic superspace:
\be
 \begin{picture}(192,10)(0,0) \put(0,0){\makebox[0pt][l]{$\bt
 \hspace {2em}
 \ominus \hspace{2em}\bt\hspace{2em}\times\hspace{2em}\bt
 \hspace{2em}\ominus \hspace{2em}\bt$}
 \rule[.5ex]{17.1em}{.1ex} }
 \end{picture}
 \ee
 This space is a super Grassmannian with local coordinates
 \be
 X^{AA'}=\left( \ba{cc} x^{\a \adt} & \l^{\a a'} \\
                        \pi^{a\adt}  &  y^{a a'}  \ea
 \right).
\la{7}
 \ee
 where $x$ are spacetime coordinates, $\l,\pi$ are odd
 coordinates and $y$ are coordinates for the internal manifold. The indices
 $(\a,\adt)$ are 2-component spacetime spinor indices while $(a,a')$ are
 2-component spinor indices for the internal space which is (locally)
 the same as spacetime in the complexified case. The capital indices
 span both spacetime and internal indices, $A=(\a,a),\ A'=(\adt,a')$,
 and we use the convention that $(\a,\adt)$ are even indices while
 $(a,a')$ are odd. As we remarked previously
 an important feature of
 analytic superspace is that superfields carrying irreducible
 representations are completely specified by the super Dynkin labels and
 analyticity; no further constraints need to be imposed.

 In the free theory the Maxwell field strength superfield,
 corresponding to the representation with $n_4=1$ and all other Dynkin
 labels zero, is a single component analytic superfield $W$. In the
 interacting case $W$ is covariantly analytic and so is not a
 superfield on analytic superspace. However, gauge-invariant
 products of $W$ are. The operators
 $A_p:=\tr(W^p)\ p=2,3,\ldots$ which transform under the representations
 which have only the central Dynkin label
 non-zero are
 in one-to-one correspondence with the Kaluza-Klein supermultiplets  of
 IIB supergravity on $AdS_5 \xz S^5$ \cite{gun,af2,hw2}. The operator
 $A_2:=T$
 is special;
 it is the supercurrent multiplet. The diagram for $A_p$ is
 \begin{picture}(75,10)(0,0)
 \put(0,0){\makebox[0pt][l]{$\bt
 \hspace {.2em}
 \ominus \hspace{.2em}\bt\hspace{.2em}\times\hspace{.2em}\bt
 \hspace{.2em}\ominus \hspace{.2em}\bt$}
 \rule[.5ex]{6.2em}{.1ex} }
 \put(36,7){\tiny p}
 \end{picture}. This means that $A_p$ is a scalar under
 $\gs\gl(2|2)\oplus\gs\gl(2|2)$ and has charge $p$ under the $U(1)$
 corresponding to the central node of the super Dynkin diagram. All other
 representations transform non-trivially
 under the sub-algebra $\gs \gl(2|2) \oplus \gs \gl(2|2)$. The series B
 superfields must transform under the totally (generalised) antisymmetric
 tensor representation (or the trivial representation) of one of the
 $\gs \gl(2|2)$ subgroups and the series C superfields must transform
 under the totally antisymmetric representation of both $\gs \gl(2|2)$
 subgroups (trivially in the KK case). For a general representation
 the highest weight state is obtained from the tensor component
 which has the most number of internal ($a$ or $a'$) indices.

We now describe how the three operators discussed earlier can be
written as fields on analytic superspace. The first one, $T_{IJ}
T_{IJ}$ in super Minkowski space, has super Dynkin labels
$(0200020)$. On analytic superspace its behaviour with respect to
both of the $\gs \gl(2|2)$ subalgebras is given by the super Dynkin
labels $(020)$. It can be constructed from two $T$'s and four
derivatives with both sets of indices, primed and unprimed, in the
representation corresponding to the super Young tableau with two
boxes in the first and second rows.

 The free Konishi multiplet on $(4,2,2)$ analytic superspace is
 \begin{picture}(75,10)(0,0)
 \put(0,0){\makebox[0pt][l]{$\bt
 \hspace {.2em}
 \ominus \hspace{.2em}\bt\hspace{.2em}\times\hspace{.2em}\bt
 \hspace{.2em}\ominus \hspace{.2em}\bt$}
 \rule[.5ex]{6.2em}{.1ex} }
 \put(13,7){\tiny 1}\put(59,7){\tiny 1}
 \end{picture}. This saturates the bounds of series A and as a tensor
 superfield has indices $K_{AB,A'B'}$ with generalised symmetry on both
 pairs. ($A$ corresponds to the left $\gs \gl(2|2)$ and $A'$ to the
 right one.) In the interacting theory, the diagram is the same with
 the 1 replaced by $1+\c$, $\c>0$. For $\c$ non-integral the
 representation \begin{picture}(40,10)(0,0)
 \put(0,0){\makebox[0pt][l]{$\bt
 \hspace {.6em}
 \ominus \hspace{.6em}\bt$}
 \rule[.5ex]{3em}{.1ex} }
 \put(12,8){\tiny 1+$\c$}
 \end{picture}
 of $\gs \gl(2|2)$ is non-tensorial; it can be explicitly described by
 putting a cross through the odd node which gives rise to a purely
 fermionic coset space of $SL(2|2)$ with four odd coordinates. The
 representation
 then has 16 components whose transformation properties can be read
 off. (If we cross both odd nodes in the full $N=4$ diagram
 we get a field on harmonic superspace which, being analytic with respect
 to the internal compact manifold, is equivalent to an
 unconstrained superfield on super Minkowski space.) The
 representations
 \begin{picture}(40,10)(0,0) \put(0,0){\makebox[0pt][l]
 {$\bt\hspace {.6em}
 \ominus \hspace{.6em}\bt$}
 \rule[.5ex]{3em}{.1ex} }
 \put(17,8){\tiny $n$}
 \end{picture}
 of $\gs \gl(2|2)$ for $n$ integral, $n \geq 2$, all have the same
 dimension as \begin{picture}(40,10)(0,0)
 \put(0,0){\makebox[0pt][l]{$\bt
 \hspace {.6em}
 \ominus \hspace{.6em}\bt$}
 \rule[.5ex]{3em}{.1ex} }
 \put(12,8){\tiny 1+$\c$}
 \end{picture}, $\c > 0$, so that the non-tensorial representation is
 closely related to these tensorial representations. In terms of the
 underlying Maxwell supermultiplet, the free Konishi superfield
 can be written \cite{hw}
 \be
 K_{AB,A'B'}=\del_{(A'(A}W\del_{B')B)}W - {1 \over 6} \del_{(A'(A
 }\del_{B')B)}W^2
\la{8}
 \ee
 However, this expression cannot be generalised to the interacting
 case since there is no gauge covariant derivative $\nabla_{\! \! A'A}$ on
 analytic superspace. Moreover \eq{8} is misleading in the quantum theory.
 The quantum Konishi multiplet resembles more closely the operator
 \begin{picture}(75,10)(0,0)
 \put(0,0){\makebox[0pt][l]{$\bt
 \hspace {.2em}
 \ominus \hspace{.2em}\bt\hspace{.2em}\times\hspace{.2em}\bt
 \hspace{.2em}\ominus \hspace{.2em}\bt$}
 \rule[.5ex]{6.2em}{.1ex} }
 \put(13,7){\tiny 2}\put(59,7){\tiny 2}
 \end{picture}
which has four primed and unprimed indices (both in the 2 times 2
box tableau) and which, by the above discussion, has the same
number of components as the interacting quantum $K$.

 We now consider the multiplet of type (iii) discussed above which is
 protected. As a field on analytic superspace it is determined by the
 diagram
 \begin{picture}(75,10)(0,0)
 \put(0,0){\makebox[0pt][l]{$\bt
 \hspace {.2em}
 \ominus \hspace{.2em}\bt\hspace{.2em}\times\hspace{.2em}\bt
 \hspace{.2em}\ominus \hspace{.2em}\bt$}
 \rule[.5ex]{6.2em}{.1ex} }
 \put(13,7){\tiny 1}\put(36,7){\tiny 2}\put(59,7){\tiny 1}
 \end{picture}. Again this
 representation has an associated anomalous representation
 \begin{picture}(75,10)(0,0)
 \put(0,0){\makebox[0pt][l]{$\bt
 \hspace {.2em}
 \ominus \hspace{.2em}\bt\hspace{.2em}\times\hspace{.2em}\bt
 \hspace{.2em}\ominus \hspace{.2em}\bt$}
 \rule[.5ex]{6.2em}{.1ex} }
 \put(8,7){\tiny 1+$\c$}\put(36,7){\tiny 2}\put(54,7){\tiny 1+$\c$}
 \end{picture} and there is also a tower of representations with $1+\c$
 replaced by
 $n\geq 2$. These all have the same dimension (as representations of
 the analytic isotropy group) as opposed to the original
 representation (with $1$'s over both the white nodes) which is smaller.
 However, as a superfield
 \begin{picture}(75,10)(0,0)
 \put(0,0){\makebox[0pt][l]{$\bt
 \hspace {.2em}
 \ominus \hspace{.2em}\bt\hspace{.2em}\times\hspace{.2em}\bt
 \hspace{.2em}\ominus \hspace{.2em}\bt$}
 \rule[.5ex]{6.2em}{.1ex} }
 \put(13,7){\tiny 1}\put(36,7){\tiny 2}\put(59,7){\tiny 1}
 \end{picture}
 can be expressed in terms of derivatives of the supercurrent $T=
 \begin{picture}(75,10)(0,0)
 \put(0,0){\makebox[0pt][l]{$\bt
 \hspace {.2em}
 \ominus \hspace{.2em}\bt\hspace{.2em}\times\hspace{.2em}\bt
 \hspace{.2em}\ominus \hspace{.2em}\bt$}
 \rule[.5ex]{6.2em}{.1ex} }
 \put(36,7){\tiny 2}
 \end{picture}$, and so in this case there is no difficulty in
 generalising the representation to the interacting case. Explicitly,
 the superfield for this representation is
 \be
 T_{AB,A'B'}=\del_{(A'(A}T \del_{B')B)}T-{1 \over 5}
 \del_{(A'(A}\del_{B')B)}T^2.
\la{10}
 \ee

We next consider operators of the form $\del^p T \del^q T$ on
analytic superspace. Since all such operators are compatible with
non-Abelian gauge invariance they are all either type (i) or type
(iii). We shall consider these operators first in the classical
theory where the Dynkin labels are all integers. Those that are
type (i) can then develop anomalous dimensions in the quantum
theory whereas the others will be protected. The result is simple:
those operators which have vanishing internal Dynkin labels are
type (i) and all the others are type (iii). This is in agreement
with the results derived in \cite{aes} using the OPE in $N=2$
harmonic superspace.

To study these operators we first define $Q=L-(J_1 + J_2)$. Since
$Q(\del)=0$ and $Q(T)=2Q(W)=2$, it follows that $Q=4$ for any of
these operators. In terms of the Dynkin labels
$Q=\sum_{i=2}^{i=6}n_i-(n_1 + n_7)$, so that we have

\be
n'_2 + n'_6 +m_1 = 4 \la{11} \ee

where $n'_2:=n_2-n_1\geq 1;\ n'_6:=n_6-n_7\geq 1$, the inequalities
following from the unitarity bounds. The requirement that the
R-charge be zero gives

\be
n_3 + 2 n_2 -n_1= n_5 + 2n_6 -n_7 \la{12} \ee

Due to the bounds we need only consider the cases $m_1=0,1,2$. For
$m_1=2$ we have $n'_2=n'_6=1$. The possible internal Dynkin labels
are $[020], [110], [011], [101], [200]$ and $[002]$.

$m_1=2; [020]$

For $[020]$ we find the super Dynkin labels are
$[k(k+1)020(k+1)k]$. These operators can be written in the form
$T\del^{k+2} T$ with the $k+2$ $A$ and $A'$ indices completely
symmetrised\footnote{Here and below we shall not write explicitly
the extra terms which are required to ensure that a given operator
is indeed primary.}. Clearly such operators saturate the bounds and
so are type (iii).

$m_1=2; [101]$

For this case the super Dynkin labels are $[k(k+1)101(k+1)k]$.
These operators can be written as $T\del^{k+3} T$ where the $(k+3)$
unprimed and primed indices are both in the representation with
symmetrisation over $(k+2)$ indices but not over all of them. Again
these operators saturate the bounds.

$m_1=2; [110]$

The super Dynkin labels are $[k(k+1)110(k+2)(k+1)]$. In this case
the left-hand $\gs\gl(2|2)$ representation corresponding to the
unprimed indices is the same as the previous case whereas the
right-hand one is totally symmetric in $k+3$ indices. These
operators cannot be written with all the derivatives hitting one of
the $T$'s, but can be written in the form $\del T\del^{k+2} T$.
Again these are saturated. The case $[011]$ is conjugate to this
one. Note that the leading component of this supermultiplet is
fermionic. In super Minkowski space it will involve an odd
derivative acting on one of the $T$'s.

$m_1=2; [200]$

Here the super Dynkin labels are $[k(k+1)200(k+3)(k+2)]$. The
left-hand $\gs\gl(2|2)$ representation has Young tableau
$<k+2,1,1>$ while the right one is $<k+4>$ where the notation
denotes the number of boxes in the first, second, third row, and so
on. It is not possible to construct this representation from
derivatives acting on two $T$'s by symmetry. The case $[002]$ is
conjugate to this one and also cannot be constructed.

$m_1=1; [010]$

If we choose $n'_2=2, n'_6=1$ we find the super Dynkin labels are
$[k(k+2)010(k+3)(k+2)]$. The corresponding tensor has left Young
tableau $<k+4>$ and right Young tableau $<k+2,2>$. These operators
can be written in the form $\del^2 T \del^{k+2} T$ and they
saturate only one of the unitarity bounds. Nevertheless, this is
sufficient for them to be of type (iii).

$m_1=1; [100]$

If we choose $n'_2=2, n'_6=1$, the super Dynkin labels are
$[k(k+2)100(k+4)(k+3)]$. The left Young tableau is $<k+2,2,1>$
while the right one is $<k+5>$. Such operators cannot be
constructed from derivatives acting on two $T$'s by symmetry.

$m_1=1; [001]$

For $n'_2=2, n'_6=1$ the Dynkin labels are $[k(k+2)001(k+2)(k+1)]$.
The left Young tableau is $<k+2,2>$ while the right one is
$<k+3,1>$. These operators can be written in the form $\del^2 T\del
^{k+2} T$ and satisfy one unitarity bound.

$m_1=0$

In this case we could in principle have $n'_2=3, n'_6=1$ but these
cannot be written in terms of derivatives acting on two $T$'s. So
take $n'_2=n'_6=2$. The super Dynkin labels are
$[k(k+2)000(k+2)k]$, and the Young tableaux are $<k+2,2>$ for both
the primed and unprimed indices. So these operators can be written
in the form $T\del^{k+4} T$ and are unsaturated. Therefore these
operators can acquire anomalous dimensions in the quantum theory.

Operators of the above form contain, as spacetime components, the
operators constructed from spacetime derivatives acting on two
factors of the leading scalars in $T$ discussed in \cite{afp1}. The
authors of \cite{afp1} were not always able to specify which
supermultiplet was involved when the component field under
discussion was not the highest weight state. Here we briefly
indicate how these supermultiplets can be identified using analytic
superspace. Let $T_o$ be the leading component of $T$; it is a
scalar field in the $20'$ representation of $SU(4)$. The operators
of \cite{afp1} are schematically of the form,

\be
\cO^{[abc]}_{r L} \sim (\del_{\adt\a})^r \square^{r'} (T_o
T_o)_{[abc]} \la{13} \ee

where $r'=1/2(L-(r+4))$, $L$ being the na\"{\i}ve dimension. The
indices on the spacetime derivatives are totally symmetrised and
$[abc]$ denotes the $SU(4)$ representation. Since $T_o$ is in the
$20'$ representation, the possible representations that can arise
are $1,20',84,105,15,175$. To illustrate the procedure let us
consider operators in the $105=[040]$ representation. There were
two series of non-renormalised $105$ operators mentioned in
\cite{afp1}, $r=2k, L= 4+2k$ and $r=2k, L= 6+2k$ (where $k$ is a
positive integer), the first non-renormalised operator being $r=0,
L=8$ . Now, as an operator on analytic superspace, the leading
component of $T^2$ is a scalar in the $105$ representation. To
obtain the desired component we therefore need only include the
right spacetime derivatives. To find the full multiplet we then
replace the spacetime derivatives by analytic superspace
derivatives.

We have $\cO^{[040]}_{2k 4+2k}\sim (\del_{\adt\a})^{2k}
(T_o^2)_{[040]}$, so the desired supermultiplet is (schematically)
$(\del_{A'A})^{2k} T^2$ with the primed and unprimed indices
symmetrised. The super Dynkin labels are
$[(2k-2)(2k-1)020(2k-1)(2k-2)]$, so this operator is protected. The
second operator is $\cO^{[040]}_{2k 6+2k}\sim
\square(\del_{\adt\a})^{2k} (T_o^2)_{[040]}$. For this case we have
$2k+2$ derivatives and the primed (unprimed) indices are
symmetrised with respect to $2k+1$ of them. In other words the
associated super Young tableaux are $<2k+1,1>$ for both sets of
indices. The super Dynkin labels are $[(2k-1)(2k)101(2k)(2k-1)]$,
and the operator is protected. In the third case $\cO^{[040]}_{0
8}\sim \square^2 (T_o^2)_{[040]}$. In this case the four
derivatives fall into the representation $<2,2>$ for both primed
and unprimed indices so the super Dynkin labels are $[0200020]$.
This  operator is unprotected.

For $k=0$, one can have no d'Alembertians, in which case the
operator is simply $T^2$, which is series C, or one can have one
d'Alembertian in which case the operator has super Dynkin labels
$[0020200]$ and is again series C.

A slightly more complicated situation arises when one needs to add
further internal derivatives in order to obtain the right $SU(4)$
representation. For example, consider the operator
$\cO^{[101]}_{2k+1\,2k+5}\sim (\del_{\adt\a})^{2k+1}
(T_o^2)_{[101]}$. Here it is necessary to add three further
derivatives. There are three possibilities corresponding to the super
Dynkin labels $[2k(2k+2)000(2k+2)2k]$ (renormalised) and
$[(2k+1)(2k+2)101(2k+2)(2k+1)]$ or $[2k(2k+2)001(2k+2)(2k+1)]$ (protected). Presumably the precise
spacetime components of the three cases will not be identical because
there will be different contributions from the other fields in the
SYM multiplet (and from terms required to make the operators
primary).

As well as the operators discussed above one can construct many
more which should be protected by the same argument. To build any
such operator one begins (schematically) with a product of $A_p$'s
and analytic superspace derivatives, with the indices on the latter
projected onto irreducible representations of the two $\gs\gl(2|2)$
superalgebras. One then requires that the operator really is
primary, i.e terms can be added in such a way to achieve this, and
finally that at least one of the unitarity bounds is satisfied.

For example,  representations with Dynkin labels
$[k(k+1)lml(k+1)k]$ can be obtained by applying derivatives to
gauge invariant operators for all positive integers $k$ and $l$ and
for all positive integers $m$ such that $m\geq 4-2l$ or $m=2-2l$.
These have the form $T\del^{k+l+2}T A_{2l+m-2}$, and since they
saturate both unitarity bounds they should be protected. Another
example is the representation $[(k+1)(k+2)lm(l+1)(k+1)k]$ for
positive integers $k,l,m$ and $m\geq 3-2l$ or $m=1-2l$. These are
of the form $\del^{k+l+2}T\del T A_{2l+m-1}$, saturate both unitarity
bounds and are therefore protected. There are also many more
examples of protected operators that saturate just one unitarity
bound.

Note that, as shown above, the only unprotected operator
constructed from two $T$'s is in the singlet representation of the
internal $SU(4)$ in agreement with \cite{aes}. Furthermore we
cannot construct any protected operators that are singlets by using
more $T$'s or $A_p$'s. There are, however, plenty of examples of
unprotected operators that are not singlets. A simple example can
be obtained by multiplying the $m_1=0$ example above by $T$. This
has the form $T^2\del^{k+4} T$ and has Dynkin labels
$[k(k+2)020(k+2)k]$ and is thus in the $20'$ representation of
$SU(4)$.

To summarise, we have seen that there are many series A composite
operators in $N=4$ SYM which should be protected from
renormalisation by virtue of the fact that they are short and
remain short in the interacting theory, whereas the corresponding
representations with anomalous dimensions are not shortened. These
protected multiplets are all multi-trace operators, since the
single-trace series A operators which are short in the free theory
do not remain short in the presence of interactions. If we write
the composite operators as fields on $(4,2,2)$ analytic superspace,
the protected operators (from any series) are analytic tensor
fields. The non-protected operators can still be interpreted as
fields on analytic superspace but they are not tensor fields of the
standard type.

{\bf Acknowledgement}

This research was supported in part by PPARC SPG grant 613.

\end{document}